\documentstyle[epsfig]{aipproc}

\begin{document}
\title{Neutrino Magnetic Moments and Atmospheric Neutrinos\thanks{
Talk presented at WIN99, Cape Town, South Africa, Jan. 25--30, 1999.\protect\\ 
This work is supported in part by KOSEF, MOE through
BSRI 98-2468, and Korea Research Foundation.
}}

\author{Jihn E. Kim}
\address{School of Physics, Korea Institute for Advanced Study,\\
207-43 Cheongryangri-dong, Seoul 130-012, Korea, and\\
Department of Physics, Seoul National University,\\
Seoul 151-742, Korea }
\maketitle
\begin{abstract}
I review the history on neutrino magnetic moments and apply the
neurino magnetic moment idea to constrain its bound from 
Super-Kamiokande neutrino oscillation data. 
\end{abstract}
\section*{Introduction}
For a long time, it was assumed that neutrinos have vanishing
quantities: $m_\nu=0, Q_{\rm em}=0$, and $\mu_\nu=0$. Among these
the neutrino mass problem has attracted the most attention, and
finally we might have an evidence for nonzero neutrino mass~\cite{Super}.
The other important neutrino property to be exploited
is the electromagnetic property, in particular the magnetc moment.

The reason for vanishing neutrino mass was very naive in 50's and
60's: the hypothesis of the $\gamma_5$-invariance. Under the 
$\gamma_5$-invariance, $\nu=\pm \gamma_5\nu$, $\nu$ appears only in one
chirality. In the standard model (SM), this is encoded as no right-handed
neutrino. 

In gauge theory models, the story changes because one can
calculate the properties of the neutrino at high precision.
In SM, one cannot write a mass term for $\nu$ in $d\le 4$ terms. To
write a mass term for $\nu$, one has to introduce $d\ge 5$ terms,
or introduce singlet neutrino(s). 
The two-component neutrino we consider in the left-handed
doublet can be Weyl or Majorana type.

If it is a Weyl neutrino, it satisfies $\nu_i=a_i\gamma_5\nu_i$
where $a_i=1$ or $-1$. Then the magnetic moment term is given by
$\bar\nu_i\sigma_{\mu\nu}q^\nu\nu_j-\bar\nu_j\sigma_{\mu\nu}q^\nu\nu_i
\rightarrow -a_ia_j[\bar\nu_i\sigma_{\mu\nu}q^\nu\nu_j-\bar
\nu_j\sigma_{\mu\nu}q^\nu\nu_i].
$
Therefore, to have a nonvanishing magnetic moment (or mass), we must
require $a_ia_j=-1$, i.e. the existence of right-handed singlet
neutrino(s).

For Majorana neutrinos, $\psi_c=C\psi^*$, it is possible to write a
mass term without introducing right-handed neutrino(s).

Thus, it is possible to introduce neutrino masses and magnetic
moments in the SM, by a slight extension of the model. The question
is how large they are.

For the detection of neutrino masses and oscillations, there have
been numerous studies from solar-, atmospheric-, reactor-,
and accelerator-neutrino experiments. The effect of neutrinos in
cosmology was also used to get bounds on neutrino masses.
On the other hand, for the neutrino magnetic moment astrophysical
constraints gave useful bounds.

Usually, the bound of the neutrino magnetic moment is given in units
($f$) of Bohr magneton ($\mu_B$),
\begin{equation}
\mu_{\nu_i}=f_i\mu_B.
\end{equation}  

\section*{History and the known bounds}

The first significant bound on magnetic moment of $\nu_e$ was
given from astrophysics, $|f_e|<10^{-10}$, by Bernstein et 
al.~\cite{bernstein}.  A better bound on $f_e$ was obtained from SN1987A, 
$|f_e|<10^{-13}$~\cite{SN}.

For the muon neutrino, the useful bound was obtained from the
neutral current data~\cite{kmo}, $|f_\mu|<0.8\times 10^{-8}$.
The first bound on the transition magnetic moment was given
also from the neutral current experiment~\cite{kim}.

For the tau neutrino, $|f_\tau|<1.3\times 10^{-7}$ has been obtained 
recently~\cite{sergei}.

For the theoretical side, it has been known from early days 
that it is possible to generate large magnetic moments for 
neutrinos~\cite{cal}. 
Note that the see-saw mass for neutrinos appear as in Fig.~1. 
Here, there does not exist any charged particle
and hence there is no contribution to magnetic moment
at this level. Thus to have a large neutrino magnetic moments,
one needs a  Feynman diagram of Fig.~2 type,
where we introduced a heavy lepton $L$ coupling to $W$
via 

\begin{figure}[b!] 
\centerline{\epsfig{file=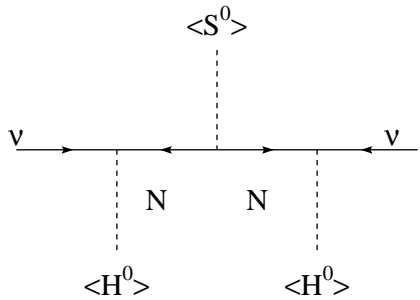,height=4cm, width=6.67cm}}
\vspace{10pt}
\caption{A see-saw mechanism for neutrino masses.}
\label{fig1}
\end{figure}

\begin{equation}
\left(\matrix{\nu_l\cr l^-\cos\alpha+L^-\sin\alpha}\right)_L\ ,\ 
\left(\matrix{L^0\cr -l^-\sin\alpha+L^-\cos\alpha}\right)_L\ ,\ 
\left(\matrix{\nu_l\cr L^-}\right)\ ,\ 
\left(\matrix{L^0\cr l^-}\right).
\end{equation}

\begin{figure}[b!] 
\centerline{\epsfig{file=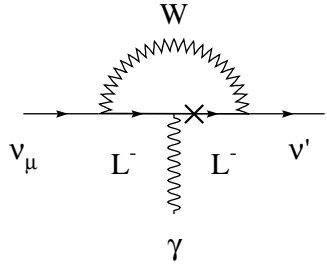,height=4cm, width=6.67cm}}
\vspace{10pt}
\caption{A Feynman diagram for neutrino magnetic moment.}
\label{fig2}
\end{figure}

Then one can easily estimate the magnetic moment of neutrinos
as~\cite{cal}\footnote{$f^\prime$ is the transition moment.}
\begin{equation}
f\ {\rm or}\ f^\prime\ =\ \frac{G_Fm_Lm_e}{2\sqrt{2}\pi^2}ab I
\left(\frac{1}{2}+\frac{1}{2}\delta_{\nu\nu^\prime}\right) 
\end{equation}
where $abI$, which is a function of mixing and Feynman integral, is 
of order 1. One can also draw Feynman diagrams with charged scalars
in the loop with appropriate Yukawa interactions introduced.
This kind of diagrams generally introduce transition magnetic
moment of order$ f^\prime \sim {m_Lm_e}/{M^2}$
where $M$ is the mass of the intermediate scalar or gauge boson.
Note that without extra charged leptons $m_L$ should be neutrino mass, 
rendering an extremely small $\mu_\nu$.

\section*{NC, $\mu^\prime_\nu$, and Single $\pi^0$ Production 
by $\nu_{\mu}^{\rm atm}$}

The Super-K collaboration has reported the ratio
\begin{equation}
R_{\pi^0/e}=\frac{(\pi^0/e)_{\rm data}}{(\pi^0/e)_{\rm MC}}
=0.93\pm 0.07\pm 0.19
\end{equation}
which is consistent with 1 at present. However, one may narrow
down the experimental errors and can observe whether it is
different from 1 or not. One assumes that $\nu_e$ is not
oscillated in the atmospheric neutrino data sample, and hence
the MC electrons are estimated with the standard CC cross section. 
The denominator is calculated assuming
that the NC is the same for the cases with and
without neutrino oscillation. So $R_{\pi^0/e}$ is expected
to be 1 if there is no oscillation of SM neutrinos to sterile 
neutrinos. If there exist oscillation of $\nu_\mu$ to sterile
neutrinos, then one expects that $R_{\pi^0/e}<1$.

However, in our recent work~\cite{kkl} we pointed out that one should
be careful to draw a firm conclusion on this matter because
if a sizable transition magnetic moment of $\nu_\mu$ exists
then one expects a different conclusion. 

For the study of NC, the single $\pi^0$ production is known
to be very useful. Most dominant contribution to the
single $\pi^0$ production at the atmospheric neutrino energy
is through $\Delta$ production,
\begin{equation}
\nu+N\rightarrow\nu+\Delta\ ,\ \ \Delta\rightarrow N+\pi^0.
\end{equation}
In this calculation, we used the form factors given in Ref.~\cite{fogli}.
For $E_\nu<10$~GeV or the kinetic energy of recoil nucleon $<1$~GeV,
the process $\nu+N\rightarrow \nu^\prime +N$ is difficult to observe at
Super-K. So the $\pi^0$ production is the cleanest way to detect NC 
interactions through Cherenkov ring (from $\pi^0$ decay) at Super-K.

For transition magnetic moment parametrized by $f^\prime$, 
$
if^\prime\mu_B\bar u(l^\prime)\sigma_{\mu\nu}q^\nu u(l)_{\nu_\mu}
$
where $q=l-l^\prime$, the single $\pi^0$ production cross section through
$\Delta$ production is given in Ref.~\cite{kkl}. In Fig.~3, we show the
result (the ratio of the neutrino magnetic moment contribution
and the NC contribution) as a function of neutrino energy.

\begin{figure}[b!] 
\centerline{\epsfig{file=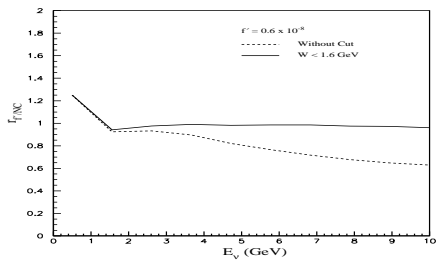,height=4cm, width=6.67cm}}
\vspace{10pt}
\caption{The ratio of $\mu^\prime_{\nu_\mu}$ and NC contributions as
a function of $E_{\nu_\mu}$. See Ref.~[7] for details.}
\label{fig3}
\end{figure}

Note that the magnetic moment part is more important at low $q^2$ region
due to the photon propagator. In principle, one can distinguish neutrino
magnetic moment interactions from the NC interactions. From Fig.~3, if
we require $r_{f^\prime/NC}\le 0.13$, then we obtain a bound $f^\prime
\le 2.2\times 10^{-9}$.

In conclusion, the transition magnetic moment $f^\prime$ can be large.
For $\nu^\prime$ heavy, it is not restricted by SN1987A bound. But 
atmospheric neutrinos of 1--10~GeV can produce $\nu^\prime$, and can 
mimick NC data~\cite{kim}. Before interpreting NC effects from atmospheric
neutrino data, one has to separate out the $\mu_{\nu}^\prime$
contribution.

\section*{Models with Large $\mu_\nu$}

Before closing, we point out $\mu_\nu$ and solar neutrino problem.
One possibility to reduce solar $\nu_e$ flux is to precess
$\nu_{eL}$ to $\nu_R$ with a large $\mu_\nu$ in a strong magnetic 
field~\cite{okun}. But this idea seems to be ruled out by
the nucleosynthesis argument~\cite{morgan}, $\mu_\nu<10^{-11}\mu_B$,
and the SN1987A argument~\cite{SN}, 
$\mu_\nu<10^{-13}\mu_B$.
The SN1987A bound is coming from
the energy loss mechanism: if $\nu_{eR}$ is created, it takes out
energy out of the core. But if it is trapped, then the bound
does not apply.\footnote{The transition magnetic 
moments to $\nu^\prime$ are not 
restricted by these bounds for a heavy enough $\nu^\prime$, 
but then the transition magnetic moment cannot account for the 
solar neutrino deficit.}

The solar neutrino problem requires $\mu_{\nu_e}\sim 10^{-11}\mu_B$
which is considered to be large. To use the idea of trapping,
the oscillation is
\begin{equation}
\nu_{eL}\rightarrow \nu^c_{\mu R}.
\end{equation}
Namely, we use the Konopinski-Mahmoud scheme where $\nu^c_{\mu R}$
is the weak interaction partner of $\mu^+_R$, implying the oscillated
neutrino participates in weak interactions; hence trapped in the
supernova core. This picture is very restrictive as suggested by
models given in Ref.~\cite{voloshin}. These models try to get a
large neutrino magnetic moment while keeping the neutrino mass
small. The $SU(2)_V$ symmetry in which $(\nu_e, \nu^c_{\mu R})$
forms a doublet under $SU(2)_V$ is introduced for this purpose.
In this case, $\mu$-number minus electron-number is conserved.
The reason for vanishing mass is that $\nu^T C\nu^c$ is symmetric under
$SU(2)_V$, i.e. the mass term is a triplet under $SU(2)_V$ and
hence is forbidden. On the other hand the magnetic moment term,
$\nu^T C\sigma_{\alpha\beta}\nu^c F^{\alpha\beta}$, is a singlet 
and is allowed, as given in Fig.~4. 
But, {\it why $SU(2)_V$?} It is like the dilemma
asked in any fermion mass ansatz problem.

\begin{figure}[b!] 
\centerline{\epsfig{file=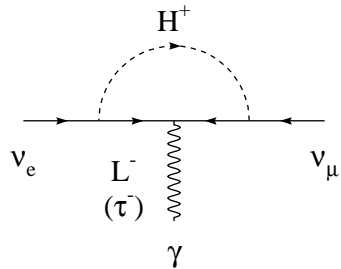,height=4cm, width=6.67cm}}
\vspace{10pt}
\caption{Neutrino magnetic moment with $SU(2)_V$.}
\label{fig4}
\end{figure}

\end{document}